\documentclass[conference]{IEEEtran}
\IEEEoverridecommandlockouts
\usepackage{cite}
\usepackage{amsmath,amssymb,amsfonts}
\usepackage{algorithmic}
\usepackage{graphicx}
\usepackage{textcomp}
\usepackage{xcolor}
\def\BibTeX{{\rm B\kern-.05em{\sc i\kern-.025em b}\kern-.08em
    T\kern-.1667em\lower.7ex\hbox{E}\kern-.125emX}}
\begin{document}

\title{Progressively Attenuated Multi-Branch Reception for Inter-HAPS Optical Links
\thanks{This work was supported by the Qatar Research Development and Innovation Council (QRDI) under Grant No. NPRP14C-0909-210008 and by research funding from Hamad Bin Khalifa University under the Thematic Research Grant Program Cycle 3. The statements made herein are solely the responsibility of the authors. The content is solely the responsibility of the authors and does not necessarily represent the official views of QRDI.

The work of M. Elamassie was supported by the Scientific and Technological Research Council of Türkiye (TÜBİTAK) under Grant 125E806.}
}

\author{
\IEEEauthorblockN{
Meysam~Ghanbari\IEEEauthorrefmark{1},
Betul~Beidas\IEEEauthorrefmark{2}
Wasiu~O.~Popoola\IEEEauthorrefmark{3},
Mazen~O.~Hasna\IEEEauthorrefmark{4},
Khalid~A.~Qaraqe\IEEEauthorrefmark{1},\\
and~Mohammed~Elamassie\IEEEauthorrefmark{2}
}

\IEEEauthorblockA{
\IEEEauthorrefmark{1}College of Science and Engineering, Hamad Bin Khalifa University, Doha, Qatar.\\
Emails: megh89467@hbku.edu.qa, kqaraqe@hbku.edu.qa
}
\IEEEauthorblockA{
\IEEEauthorrefmark{2}Department of Electrical and Electronics Engineering, \"{O}zye\v{g}in University, Istanbul, Türkiye.\\
Emails: betul.beidas@ozu.edu.tr, mohammed.elamassie@ozyegin.edu.tr
}
\IEEEauthorblockA{
\IEEEauthorrefmark{3}Electronics and Electrical Engineering, The University of Edinburgh, Edinburgh, United Kingdom.\\
Email: W.Popoola@ed.ac.uk
}

\IEEEauthorblockA{
\IEEEauthorrefmark{4}Department of Electrical Engineering, Qatar University, Doha, Qatar.\\
Email: hasna@qu.edu.qa
}
}

\maketitle

\begin{abstract}
Inter-HAPS optical links can experience receiver saturation at short separations and become signal-to-noise-ratio (SNR) limited at longer distances. This work proposes a saturation-aware spatial multi-branch intensity-modulation/direct-detection receiver using nonoverlapping apertures with progressive attenuation and selection of the nonsaturated branch with the highest instantaneous SNR. A Gaussian-beam model with finite-aperture pointing loss and common two-dimensional pointing jitter yields closed-form branch-level expressions for saturation, insufficient-SNR, usability, and outage probabilities. Results show that additional branches suppress short-range saturation outage, shift the minimum-outage point toward shorter separations, and enable progressive attenuation to control the saturation–SNR tradeoff. Monte Carlo simulations closely match the analytical results. Overall, the proposed architecture provides a low-complexity approach to extending the usable dynamic range of inter-HAPS optical receivers.
\end{abstract}


\section{Introduction}
High-altitude platform stations (HAPS) are emerging as a promising component of future non-terrestrial networks, providing wide-area coverage and flexible connectivity from the stratosphere. For inter-HAPS links, free-space optical (FSO) communication is particularly attractive because it offers very high data rates, large available bandwidth, narrow-beam transmission, and immunity to radio-frequency spectrum congestion. These advantages make optical links well suited for high-capacity HAPS backhaul and inter-platform connectivity \cite{Ghanbari2024Beamwidth}.
Despite these benefits, inter-HAPS optical links can experience large variations in received optical power due to changes in link distance, Gaussian-beam spreading, and residual pointing error. At short separations or under strong beam coupling, the received power may exceed the linear operating range of the photodetector or front end, resulting in receiver saturation. At longer distances, however, reduced optical coupling can drive the link into an SNR-limited regime. Therefore, maintaining reliable reception over a wide operating range requires simultaneously addressing both high-power saturation and low-power SNR degradation \cite{Huang2023SPAD,Ghanbari2026NarrowBeams}.

Several approaches have been investigated to improve the robustness of 
FSO links under such channel and power variations. HAPS-based optical 
links have been analyzed under attenuation, pointing error, 
angle-of-arrival fluctuations, atmospheric turbulence, and 
adaptive-optics correction \cite{Ata2023HAPS}. Spatial-diversity reception using 
multiple apertures has also been investigated to improve receiver 
sensitivity and BER under turbulent conditions \cite{Huang2022SpatialDiversity}. At the transmitter 
side, adaptive optical-power regulation has been experimentally 
demonstrated to compensate for large variations in received power \cite{Brandao2024FSO}.

Receiver-side approaches have also been developed to accommodate a wide 
optical dynamic range. A hybrid SPAD/PD receiver combining detector-mode 
switching with variable optical attenuation was proposed to support 
reliable operation under both low- and high-power conditions \cite{Huang2020Hybrid}, and 
its ability to extend the receiver operating range was subsequently 
demonstrated experimentally \cite{Huang2021Reliable}. Automatic attenuation control has also 
been employed to mitigate photon-counting saturation by dynamically 
adjusting the optical power incident on SPAD-based receivers \cite{Wang2022BER,Wang2026Automatic}. 
Earlier work on multi-element CMOS imaging receivers also considered 
wide-dynamic-range optical reception for FSO systems \cite{Leibowitz2005CMOS}.

In contrast to these approaches, the receiver proposed in this work does 
not rely on detector-mode switching, dynamically controlled attenuation, 
or a conventional multi-element imaging architecture. Instead, it 
combines spatially separated receive apertures with deliberately unequal 
fixed pre-detection attenuation levels. The resulting branches provide 
complementary operating regions: strongly attenuated branches remain 
usable under high received-power conditions where saturation dominates, 
whereas lightly attenuated branches retain higher SNR as the link moves 
toward the low-power regime. Saturation-aware branch selection then 
exploits these complementary operating regions by selecting the 
highest-SNR branch among those that remain within the receiver linear 
operating region.

The main contributions of this work are threefold. First, we propose a saturation-aware spatial multi-branch IM/DD receiver that combines physically separated apertures with progressively stronger fixed attenuation, creating complementary operating regions across saturation- and SNR-limited conditions. Second, we derive closed-form branch-level expressions for the saturation, insufficient-SNR, usability, and outage probabilities under finite-aperture Gaussian-beam coupling and two-dimensional pointing jitter. Third, we formulate the exact system-level outage probability while preserving the common pointing realization and resulting cross-branch dependence. Analytical and Monte Carlo results demonstrate substantial suppression of short-range saturation outage and an extended usable dynamic range for inter-HAPS optical links.

\section{System Model and Spatial Multi-Branch Receiver}
\label{sec:system_model}

\begin{figure}[t]
    \centering
    \includegraphics[width=\columnwidth]{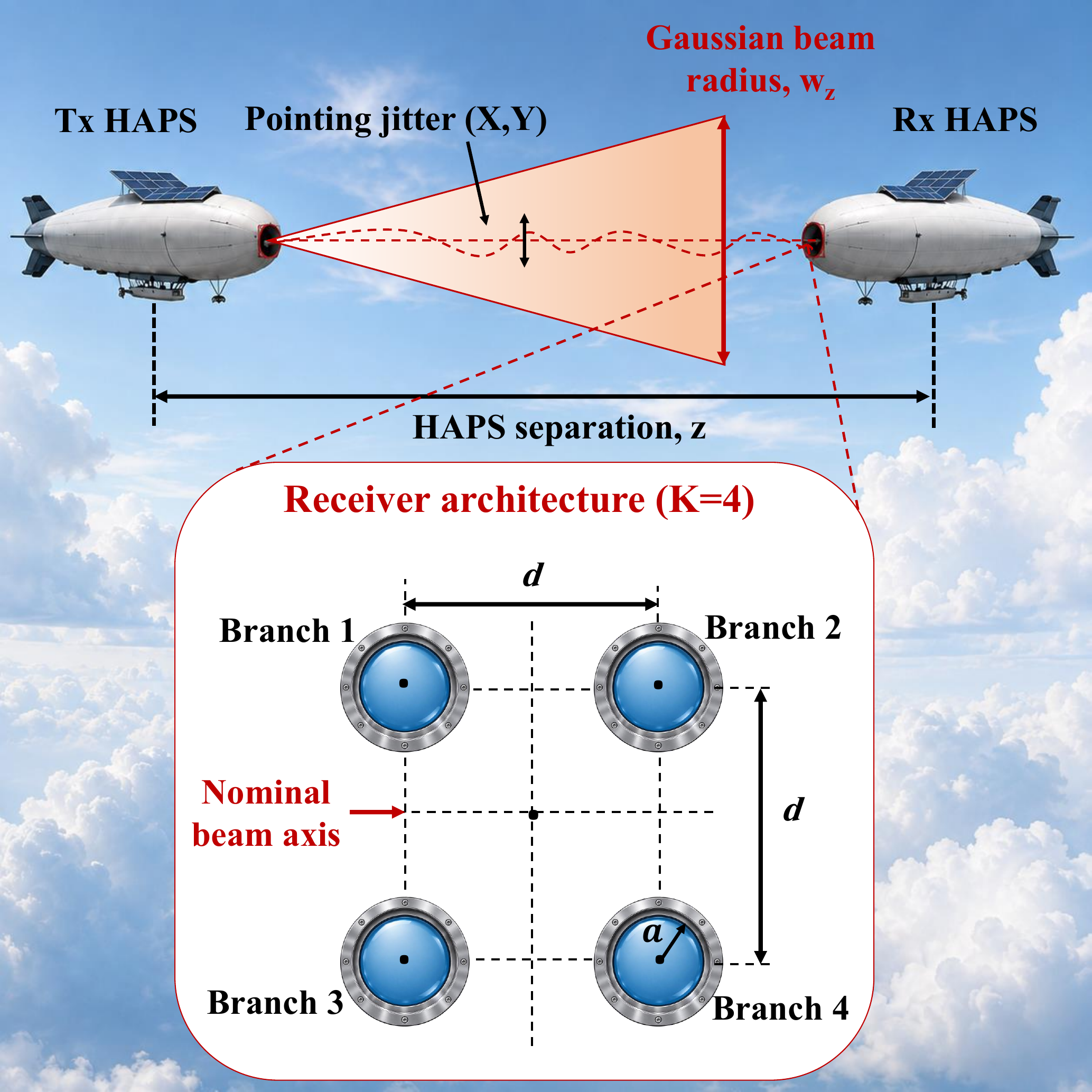}
    \caption{Inter-HAPS optical link and $(K=4)$ spatial multi-branch receiver architecture with pointing displacement and receiver geometry.}
    \label{fig:fig1}
\end{figure}

We consider a line-of-sight inter-HAPS optical link employing intensity modulation and direct detection
(IM/DD). The terminals operate in the stratosphere under clear-air conditions, where turbulence-induced
scintillation is typically negligible and is therefore omitted from the adopted model. The model accounts for
Gaussian-beam spreading, residual pointing error, finite-aperture spatial collection, optical loading, and
receiver saturation. As illustrated in Fig.~\ref{fig:fig1}, the Gaussian beam propagates over the HAPS separation $z$, while
residual pointing jitter produces a common beam-center displacement $(X,Y)$ at the spatial multi-branch
receiver. The $K=4$ square receiver configuration is also illustrated, with aperture radius $a$ and center-to-
center spacing $d$.
We model the instantaneous transmitted optical power as
\begin{equation}
P_t(t)
=
P_{\mathrm{DC}} + P_s x(t),
\label{eq:transmitted_power}
\end{equation}
where $P_t(t)$ is the instantaneous transmitted optical power, $P_{\mathrm{DC}}$ denotes the optical DC bias, $P_s$ scales the information-bearing optical component, and $x(t)$ is a normalized zero-mean modulation waveform satisfying $|x(t)| \leq 1$. Its normalized average signal power is defined as $P_x \triangleq \mathbb{E}[x^2(t)]$.
The corresponding
peak transmitted optical power is 
\begin{equation}
P_{t,\mathrm{pk}}
=
P_{\mathrm{DC}} + P_s,
\label{eq:peak_transmitted_power}
\end{equation}
where $P_{t,\mathrm{pk}}$ denotes the peak transmitted optical power.

\subsection{Gaussian Beam and Residual Pointing Model}
\label{subsec:gaussian_pointing}
Let $z$ denote the inter-HAPS separation, $\lambda$ the optical wavelength, and
$w_0$ the $1/e^2$-intensity beam-waist radius. The Rayleigh range is
$z_R=\pi w_0^2/\lambda$, and the Gaussian-beam radius at distance $z$ is
$w_z=w_0\sqrt{1+(z/z_R)^2}$ \cite{Ghanbari2026NarrowBeams}.
Each branch employs a circular optical collection aperture of radius $a$. 
Using the equivalent-beam finite-aperture model \cite{Ghanbari2026NarrowBeams}, we define
$v(z)=\sqrt{\pi}a/(\sqrt{2}w_z)$,
$A_0(z)=\operatorname{erf}^2[v(z)]$, and
$w_{\mathrm{eq}}^2(z)=
w_z^2\sqrt{\pi}\operatorname{erf}[v(z)]/
\left(2v(z)e^{-v^2(z)}\right)$.
Here, $v(z)$ is the aperture-to-beam parameter, $\operatorname{erf}(\cdot)$ denotes the error function, $A_0$ is the maximum finite-
aperture coupling coefficient at zero relative displacement, and $w_{\mathrm{eq}}$ is the corresponding equivalent beam
radius.
The residual pointing errors are modeled as independent Gaussian variables,
$\theta_x,\theta_y \sim \mathcal{N}(0,\sigma_\theta^2)$.
Using the small-angle approximation, the corresponding receiver-plane
displacements are $X=z\theta_x$ and $Y=z\theta_y$, with per-axis standard
deviation $\sigma_s=z\sigma_\theta$.
The receiver comprises \(K \in \{1,\ldots,5\}\) nonoverlapping circular collection elements. Denoting the center of branch \(k\) by \((x_k,y_k)\), the branch-center set is defined as
\begin{equation}
\mathcal{C}_K \triangleq \left\{(x_k,y_k)\right\}_{k=1}^{K}.
\end{equation}

The adopted symmetric geometries are a centered aperture for \(K=1\); a pair at \((\pm d/2,0)\) for \(K=2\); three collinear apertures at \(x_k \in \{-d,0,d\}\) for \(K=3\); a square with coordinates \((\pm d/2,\pm d/2)\) for \(K=4\); and a central aperture with four cardinal apertures at \((\pm d,0)\) and \((0,\pm d)\) for \(K=5\). Here, \(d\) denotes the adjacent center spacing, and \(d \geq 2a\) ensures nonoverlapping apertures.
The receiver is a spatial array rather than a beam-splitting architecture.
Consequently, all branches experience the same instantaneous beam-center
realization $(X,Y)$, while their relative displacements and spatial coupling
coefficients differ according to their positions. For branch $k$, the
beam-center displacement relative to its center is
$r_k=\sqrt{(X-x_k)^2+(Y-y_k)^2}$, and the corresponding finite-aperture
pointing-coupling coefficient is

\begin{equation}
h_{p,k}
=
A_0\exp\left(-\frac{2r_k^2}{w_{\mathrm{eq}}^2}\right).
\label{eq:branch_pointing_coupling}
\end{equation}

Hence, $0<h_{p,k}\leq A_0$, and branches at different spatial locations
generally experience distinct $h_{p,k}$ under the same pointing realization.
For statistical characterization, we define the radial offset of branch $k$
from the nominal optical axis as $\nu_k=\sqrt{x_k^2+y_k^2}$.
The displacement $r_k$ then follows a Rician distribution with CDF \cite{Ismail2018FSO}
\begin{equation}
F_{r_k}(r)
=
1-Q_1\left(
\frac{\nu_k}{\sigma_s},
\frac{r}{\sigma_s}
\right),
\qquad
r\geq 0,
\label{eq:rician_cdf}
\end{equation}
where $Q_1(\cdot,\cdot)$ is the first-order Marcum-$Q$ function. For a centered branch, $\nu_k=0$, and
\eqref{eq:rician_cdf} reduces to the Rayleigh distribution. Since all $r_k$ are generated from the common random pair
$(X,Y)$, the branch pointing gains are generally correlated; therefore, \eqref{eq:rician_cdf} characterizes only their marginal distributions.

\subsection{Spatially Attenuated Photodetection Branches}
\label{subsec:attenuated_branches}

Each branch comprises a finite circular collection aperture, an optical attenuation element with power
transmission coefficient $\alpha_k$, an identical photodetector/front end, and a subsequent DC-blocking stage. The
attenuation coefficients satisfy
\begin{equation}
\begin{aligned}
1=\alpha_1>\alpha_2>\cdots>\alpha_K>0,
\end{aligned}
\label{eq:attenuation_progression}
\end{equation}
The attenuation coefficients follow the progression
$\alpha_k=\rho^{k-1}$ with $0<\rho<1$, where $\rho$ is the
attenuation-progression parameter.
This geometric progression is adopted because it provides a monotonic set of attenuation levels using a single design parameter, while maintaining a constant attenuation ratio between adjacent branches. Consequently, $\rho$ directly controls the tradeoff between saturation protection and SNR preservation without requiring independent optimization of all $K$ attenuation coefficients.
Let $h_l(z)$ denote the common deterministic optical efficiency. 
Denoting by $P_B$ the aggregate background optical power collected by an equal receiving aperture before attenuation, the desired collected optical power in branch $k$ is
\begin{equation}
P_{\mathrm{sig},k}(t)
=
h_l h_{p,k}
\left[P_{\mathrm{DC}}+P_s x(t)\right].
\label{eq:branch_signal_power}
\end{equation}
After branch-specific attenuation, the total photodetector-incident optical power is
\begin{equation}
P_k(t)
=
\alpha_k
\left\{
h_l h_{p,k}
\left[P_{\mathrm{DC}}+P_s x(t)\right]
+
P_B
\right\}.
\label{eq:branch_total_power}
\end{equation}

We assume spatially uniform background radiance over the compact array, such that the same $P_B$ applies to
all branches before branch-specific attenuation. Since the attenuator precedes photodetection, $\alpha_k$ scales the
desired DC component, the information-bearing optical component, and the background optical power.
For identical photodetectors with responsivity $R$, the pre-DC-block photocurrent in branch $k$ is
\begin{equation}
i_k(t)
=
R\alpha_k
\left\{
h_l h_{p,k}\left[P_{\mathrm{DC}}+P_s x(t)\right]
+
P_B
\right\}.
\label{eq:branch_photocurrent}
\end{equation}

Let $I_{\mathrm{sat}}$ denote the effective photocurrent threshold beyond which the photodetector or associated analog
front end is regarded as nonlinear. Since $|x(t)|\leq 1$, the maximum branch current for a given pointing
realization is

\begin{equation}
\begin{aligned}
i_{k,\max}
&=
R\alpha_k
\left[
h_l h_{p,k}\left(P_{\mathrm{DC}}+P_s\right)+P_B
\right],
\\
\text{linear branch:}\qquad
i_{k,\max}
&\leq I_{\mathrm{sat}}.
\end{aligned}
\label{eq:branch_saturation}
\end{equation}

The DC-blocking stage follows the photodetector/front end and therefore does not alter the optical loading
that determines the saturation condition in \eqref{eq:branch_saturation}. For a branch operating within its linear region, DC blocking
removes the mean electrical photocurrent, yielding the information-bearing current
\begin{equation}
i_{s,k}(t)
=
R\alpha_k h_l h_{p,k} P_s x(t),
\label{eq:useful_current}
\end{equation}
where $i_{s,k}(t)$ denotes the DC-blocked useful current. Accordingly, $P_{\mathrm{DC}}$ does not contribute to the useful AC
current in \eqref{eq:useful_current}, although it remains part of the saturation-relevant optical loading in \eqref{eq:branch_saturation}.
We adopt a background-shot-noise-limited receiver model. The background photocurrent in branch $k$ is
$I_{B,k}=R\alpha_k P_B$. Hence, denoting the one-sided equivalent electrical noise bandwidth by $B$, the corresponding integrated noise variance $\sigma_{b,k}^2$ is \cite{Wang2016BER}
\begin{equation}
\sigma_{b,k}^2
=
2qR\alpha_k P_B B,
\label{eq:background_shot_noise}
\end{equation}
where $q$ denotes the elementary charge and $f$ is electrical frequency. Although DC blocking removes the
mean background photocurrent from the information-detection path, the associated shot-noise fluctuations
remain. Other noise sources, including thermal noise, dark-current noise, desired-signal shot noise, relative-
intensity noise, quantization noise, and amplifier noise, are excluded from the adopted baseline model.

\section{Saturation-Aware Receiver Selection and Performance Analysis}
\label{sec:selection_performance}

For a branch operating in its linear region, the DC-blocked information current is given by
\eqref{eq:useful_current}, whereas the background-shot-noise variance is given by
\eqref{eq:background_shot_noise}. For a given pointing realization, the electrical SNR of
branch $k$ is therefore

\begin{equation}
\gamma_k
=
\frac{
R^2\alpha_k^2 h_l^2 h_{p,k}^2 P_s^2 P_x
}{
2qR\alpha_k P_B B
}
=
\frac{
R\alpha_k h_l^2 h_{p,k}^2 P_s^2 P_x
}{
2qP_B B
}.
\label{eq:branch_snr}
\end{equation}

Thus, the branch SNR depends jointly on its attenuation coefficient and instantaneous spatial coupling
through $\alpha_k h_{p,k}^2$.
For a given realization of $(X,Y)$, we define the set of branches satisfying the linear-region constraint in
\eqref{eq:branch_saturation} as

\begin{equation}
\mathcal{F}=\left\{k:\; i_{k,\max}\le I_{\mathrm{sat}}\right\}.
\label{eq:feasible_set}
\end{equation}

When $\mathcal{F}\neq\varnothing$, the receiver selects

\begin{equation}
k^\star
=
\underset{k\in\mathcal{F}}{\arg\max}\;
\gamma_k.
\label{eq:branch_selection}
\end{equation}

Since the factor $Rh_l^2P_s^2P_x/(2qP_BB)$ in \eqref{eq:branch_snr} is common and positive,
\eqref{eq:branch_selection} is equivalently an optimization of $\alpha_k h_{p,k}^2$ over the feasible branches.
Consequently, ordering the branches according only to $\alpha_k$ is not generally optimal because the spatial
array produces branch-dependent $h_{p,k}$.
For a prescribed minimum acceptable SNR $\gamma_{\min}>0$, the instantaneous receiver state is classified as
\begin{equation}
\mathcal{S}
=
\begin{cases}
\text{saturation outage}, 
& \mathcal{F}=\varnothing,\\[2mm]
\text{insufficient-SNR outage}, 
& \mathcal{F}\neq\varnothing,\quad
\displaystyle \max_{k\in\mathcal{F}}\gamma_k<\gamma_{\min},\\[2mm]
\text{valid operation}, 
& \mathcal{F}\neq\varnothing,\quad
\displaystyle \max_{k\in\mathcal{F}}\gamma_k\geq\gamma_{\min}.
\end{cases}
\label{eq:receiver_state}
\end{equation}

In the valid state, the detected signal is obtained from branch $k^\star$.

\subsection{Saturation and SNR Boundaries}
\label{subsec:sat_snr_boundaries}

Using $P_{t,\mathrm{pk}}=P_{\mathrm{DC}}+P_s$, the linearity condition in
\eqref{eq:branch_saturation} can be expressed as
\begin{equation}
h_{p,k}\leq h_{\mathrm{sat},k},
\qquad
h_{\mathrm{sat},k}
\triangleq
\frac{I_{\mathrm{sat}}/(R\alpha_k)-P_B}
{h_l P_{t,\mathrm{pk}}}.
\label{eq:saturation_coupling_threshold}
\end{equation}

The physical range $0<h_{p,k}\leq A_0$ leads to three distinct saturation regimes. If $R\alpha_k P_B\geq I_{\mathrm{sat}}$,
background loading alone reaches the saturation threshold, and branch $k$ is saturated for every finite
displacement. If $I_{\mathrm{sat}}
\geq
R\alpha_k
\left[
h_l A_0 P_{t,\mathrm{pk}}+P_B
\right]$, the branch remains linear even at maximum coupling and therefore never saturates. In the intermediate
regime, $R\alpha_k P_B
<
I_{\mathrm{sat}}
<
R\alpha_k
\left[
h_l A_0 P_{t,\mathrm{pk}}+P_B
\right]$, we have $0<h_{\mathrm{sat},k}<A_0$. Substitution of \eqref{eq:branch_pointing_coupling} into
\eqref{eq:saturation_coupling_threshold} then yields the saturation displacement boundary
\begin{equation}
r_{\mathrm{sat},k}
=
\frac{w_{\mathrm{eq}}}{\sqrt{2}}
\sqrt{
\ln\left(
\frac{A_0}{h_{\mathrm{sat},k}}
\right)
}.
\label{eq:saturation_displacement_boundary}
\end{equation}

In this intermediate regime, branch $k$ is saturated for $r_k<r_{\mathrm{sat},k}$ and linear for
$r_k\geq r_{\mathrm{sat},k}$.
Using the unified Rician displacement model in \eqref{eq:rician_cdf}, the marginal saturation probability of branch $k$ is
\begin{equation}
P_{\mathrm{sat},k}
=
\left\{
\begin{array}{@{}l@{\hspace{4pt}}l@{}}
1,
& R\alpha_k P_B \geq I_{\mathrm{sat}},\\[2mm]

0,
& I_{\mathrm{sat}}
\geq
R\alpha_k
\left[
h_l A_0 P_{t,\mathrm{pk}}+P_B
\right],\\[2mm]

1-Q_1\!\left(
\frac{\nu_k}{\sigma_s},
\frac{r_{\mathrm{sat},k}}{\sigma_s}
\right),
& \text{otherwise}.
\end{array}
\right.
\label{eq:saturation_probability}
\end{equation}

The corresponding nonsaturation probability is
\begin{equation}
P_{\mathrm{ns},k}
=
1-P_{\mathrm{sat},k}.
\label{eq:nonsaturation_probability}
\end{equation}

We next impose the minimum-SNR requirement. From \eqref{eq:branch_snr}, $\gamma_k\geq\gamma_{\min}$ is equivalent to
\begin{equation}
h_{p,k}
\geq
h_{\mathrm{SNR},k},
\qquad
h_{\mathrm{SNR},k}
\triangleq
\sqrt{
\frac{
2qP_BB\gamma_{\min}
}{
R\alpha_k h_l^2 P_s^2 \sigma_x^2
}
}.
\label{eq:snr_coupling_threshold}
\end{equation}

When $h_{\mathrm{SNR},k}<A_0$, substitution of \eqref{eq:branch_pointing_coupling} into
\eqref{eq:snr_coupling_threshold} gives the maximum displacement satisfying the SNR
requirement,
\begin{equation}
r_{\mathrm{SNR},k}
=
\frac{w_{\mathrm{eq}}}{\sqrt{2}}
\sqrt{
\ln\left(
\frac{A_0}{h_{\mathrm{SNR},k}}
\right)
}.
\label{eq:snr_displacement_boundary}
\end{equation}

Hence, $\gamma_k\geq\gamma_{\min}$ for $r_k\leq r_{\mathrm{SNR},k}$. If
$h_{\mathrm{SNR},k}\geq A_0$, the branch does not provide a positive-probability
region satisfying the required SNR under the continuous pointing model.
For branches that are not always saturated, we define the lower displacement boundary of their linear region
as
\begin{equation}
r_{\mathrm{L},k}
=
\begin{cases}
0,
& I_{\mathrm{sat}}
\geq
R\alpha_k
\left[
h_l A_0 P_{t,\mathrm{pk}}+P_B
\right],\\[2mm]
r_{\mathrm{sat},k},
& R\alpha_k P_B
<
I_{\mathrm{sat}}
<
R\alpha_k
\left[
h_l A_0 P_{t,\mathrm{pk}}+P_B
\right].
\end{cases}
\label{eq:lower_displacement_boundary}
\end{equation}

Thus, a usable branch requires the simultaneous displacement conditions $
r_{\mathrm{L},k}
\leq
r_k
\leq
r_{\mathrm{SNR},k}$, where saturation excludes sufficiently small
branch-relative displacements, whereas insufficient SNR excludes sufficiently large displacements.

\subsection{Closed-Form Branch-Level Probabilities}
\label{subsec:branch_probabilities}

We define branch-level insufficient-SNR operation as the event that branch $k$ is nonsaturated but does not
meet $\gamma_{\min}$. For a branch that is not always saturated, its probability is
\begin{equation}
P_{\mathrm{IS},k}
=
\begin{cases}
P_{\mathrm{ns},k},
& h_{\mathrm{SNR},k}\geq A_0,\\[2mm]
Q_1\left(
\displaystyle \frac{\nu_k}{\sigma_s},
\displaystyle
\frac{\max\{r_{\mathrm{L},k},r_{\mathrm{SNR},k}\}}{\sigma_s}
\right),
& h_{\mathrm{SNR},k}<A_0.
\end{cases}
\label{eq:insufficient_snr_probability}
\end{equation}

For an always-saturated branch, $P_{\mathrm{IS},k}=0$.
A branch is usable when it is simultaneously nonsaturated and satisfies the SNR constraint. Its marginal
usable probability is therefore
\begin{equation}
P_{\mathrm{use},k}
=
\begin{cases}
0,
& R\alpha_k P_B\geq I_{\mathrm{sat}},\\[2mm]
0,
& h_{\mathrm{SNR},k}\geq A_0,\\[2mm]
0,
& r_{\mathrm{L},k}\geq r_{\mathrm{SNR},k},\\[2mm]
\begin{aligned}
&Q_1\left(
\displaystyle \frac{\nu_k}{\sigma_s},
\displaystyle \frac{r_{\mathrm{L},k}}{\sigma_s}
\right)
\\[-1mm]
&\quad -
Q_1\left(
\displaystyle \frac{\nu_k}{\sigma_s},
\displaystyle \frac{r_{\mathrm{SNR},k}}{\sigma_s}
\right),
\end{aligned}
& \text{otherwise}.
\end{cases}
\label{eq:usable_probability}
\end{equation}

For every branch, the decomposition
$
P_{\mathrm{ns},k}
=
P_{\mathrm{IS},k}
+
P_{\mathrm{use},k},
$
follows directly from the mutually exclusive nonsaturated operating states.
The total marginal outage probability of branch $k$ is the probability that the branch is either saturated or
nonsaturated but fails to satisfy the SNR requirement. Since these two events are mutually exclusive,
\begin{equation}
P_{\mathrm{out},k}
=
P_{\mathrm{sat},k}
+
P_{\mathrm{IS},k}
=
1-P_{\mathrm{use},k}.
\label{eq:branch_outage_probability}
\end{equation}

Thus, the branch-level outage probability is available in closed form through
\eqref{eq:saturation_probability}, \eqref{eq:insufficient_snr_probability}, and
\eqref{eq:usable_probability}.
For a centered branch, $\nu_k=0$ and $Q_1(0,b)=\exp(-b^2/2)$. The intermediate-regime saturation
probability and usable probability consequently reduce to

\begin{equation}
P_{\mathrm{sat},k}
=
1-
\exp\left(
-\frac{r_{\mathrm{sat},k}^2}{2\sigma_s^2}
\right).
\label{eq:centered_branch_probabilities}
\end{equation}
Similarly, the usable probability becomes
\begin{equation}
P_{\mathrm{use},k}
=
\exp\left(
-\frac{r_{\mathrm{L},k}^2}{2\sigma_s^2}
\right)
-
\exp\left(
-\frac{r_{\mathrm{SNR},k}^2}{2\sigma_s^2}
\right).
\label{eq:centered_branch_use_probability}
\end{equation}

Whenever the corresponding finite usable interval exists. For off-axis branches, the Marcum-$Q$ forms in
\eqref{eq:saturation_probability}, \eqref{eq:insufficient_snr_probability}, and
\eqref{eq:usable_probability} retain the dependence on the normalized spatial offset $\nu_k/\sigma_s$.

\subsection{Exact System-Level Outage Probability}
\label{subsec:system_outage}

The quantities in \eqref{eq:saturation_probability}--\eqref{eq:centered_branch_use_probability} are marginal branch probabilities and cannot be multiplied across branches to
obtain system-level outage probabilities because all $h_{p,k}$ depend on the same random pair $(X,Y)$. Let
$\mathcal{R}_c\subset\mathbb{R}^2$ denote the region in the common pointing plane associated with any receiver state $c$ defined in
\eqref{eq:receiver_state}. Its exact probability is
\begin{equation}
P_c
=
\frac{1}{2\pi\sigma_s^2}
\iint_{\mathcal{R}_c}
\exp\left[
-\frac{X^2+Y^2}{2\sigma_s^2}
\right]
\,dX\,dY.
\label{eq:state_region_probability}
\end{equation}

\begin{figure}[t]
    \centering
    \includegraphics[width=\columnwidth]{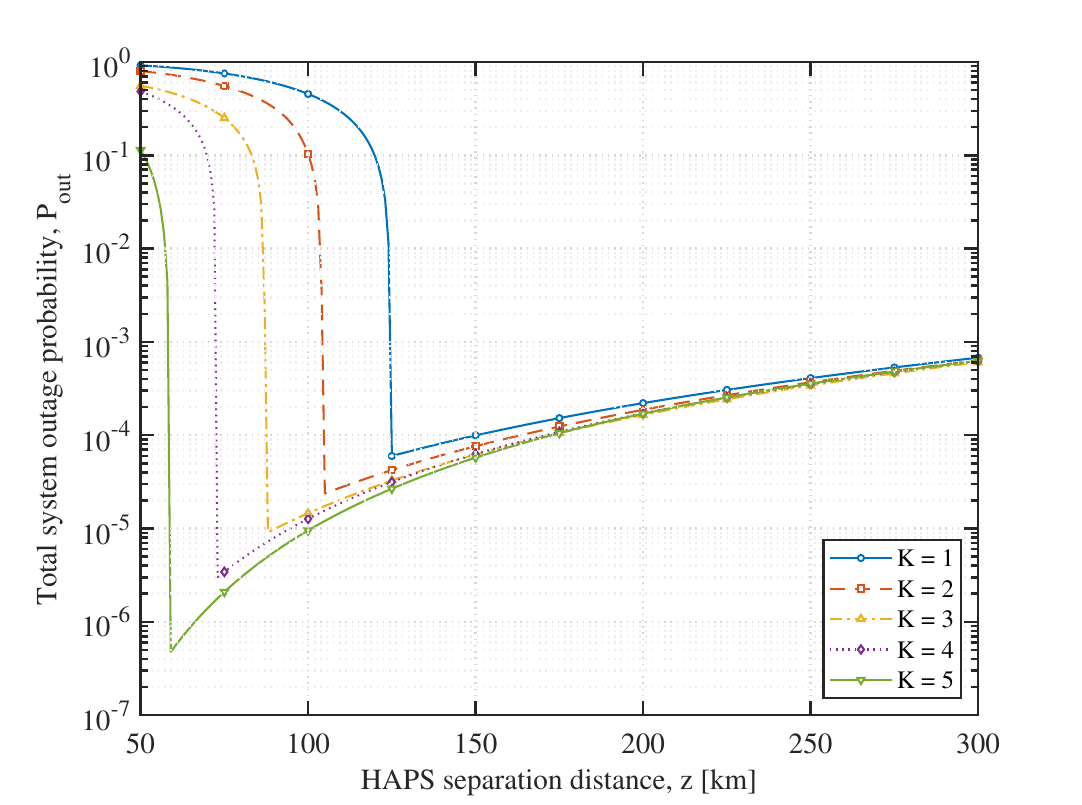}
\caption{Total system outage probability $P_{\mathrm{out}}$ versus HAPS separation distance $z$ for receiver configurations $K=1,\ldots,5$, with pointing-jitter standard deviation $\sigma_{\theta}=10~\mu\mathrm{rad}$ and saturation-current threshold $I_{\mathrm{sat}}=5~\mathrm{mA}$.}
    \label{fig:fig2}
\end{figure}

Because the boundaries of $\mathcal{R}_c$ are formed jointly by several displaced branch-centered saturation and SNR
regions, \eqref{eq:state_region_probability} does not, in general, factor into marginal branch probabilities. The branch-level closed forms
above therefore provide exact marginal statistics, while system-level performance retains the full cross-
branch pointing dependence through the common two-dimensional pointing realization.
Let $P_{\mathrm{sat,out}}$, $P_{\mathrm{SNR,out}}$, and $P_{\mathrm{val}}$ denote \eqref{eq:state_region_probability} evaluated over the saturation-outage, insufficient-SNR-outage,
and valid-operation regions defined in \eqref{eq:receiver_state}, respectively. Since these three receiver states are mutually
exclusive and exhaustive, the total system outage probability is
\begin{equation}
P_{\mathrm{out}}
=
P_{\mathrm{sat,out}}
+
P_{\mathrm{SNR,out}}
=
1-P_{\mathrm{val}}.
\label{eq:system_outage_probability}
\end{equation}

whereas
\[
P_{\mathrm{SNR,out}}
=
\Pr\left\{
\mathcal{F}\neq\varnothing,\;
\max_{k\in\mathcal{F}}\gamma_k<\gamma_{\min}
\right\}.
\]

Hence, $P_{\mathrm{out}}$ jointly accounts for receiver saturation and insufficient received SNR while retaining the full
cross-branch dependence induced by the common pointing realization.

\section{Simulation Results}
\label{sec:simulation_results}

In this section, the analytical results are evaluated numerically and validated through Monte Carlo
simulations. Unless otherwise stated, the system parameters are set to $\lambda=1550$ nm, $w_0=0.021$ m,
$P_{\mathrm{DC}}=4$ W, $P_s=0.2$ W, $h_l=0.5$, $a=0.10$ m, $d=0.45$ m, $R=0.85$ A/W,
$I_{\mathrm{sat}}=5$ mA, $P_B=50$ nW, $B=10$ GHz, $\sigma_\theta=10~\mu$rad, and
$\gamma_{\min}=25$ dB \cite{Ghanbari2024Beamwidth}. Monte Carlo results use $2\times10^5$ realizations of the
common two-dimensional pointing displacement $(X,Y)$.

Fig.~\ref{fig:fig2} shows the total system outage probability versus HAPS separation for
$K=1,\ldots,5$. At short distances, outage is dominated by receiver saturation,
and the additional progressively attenuated branches provide nonsaturated
alternatives; at $z=50~\mathrm{km}$, $P_{\mathrm{out}}$ decreases from
approximately $0.92$ for $K=1$ to $0.11$ for $K=5$. Increasing $K$ also shifts
the minimum-outage region toward shorter distances and enables very low outage
levels, reaching the $10^{-5}$--$10^{-6}$ range. For example, at
$z=100~\mathrm{km}$, $P_{\mathrm{out}}$ is approximately
$1.46\times10^{-5}$, $1.27\times10^{-5}$, and $9.53\times10^{-6}$ for
$K=3$, $4$, and $5$, respectively. Beyond the minimum, outage gradually
increases as beam spreading and pointing displacement reduce the received SNR.
Although the curves become closer at longer distances, the multi-branch
receivers generally retain a modest advantage over the single-aperture case
because the spatially displaced apertures provide additional opportunities for
favorable beam coupling under pointing jitter.

\begin{figure}[t]
    \centering
    \includegraphics[width=\columnwidth]{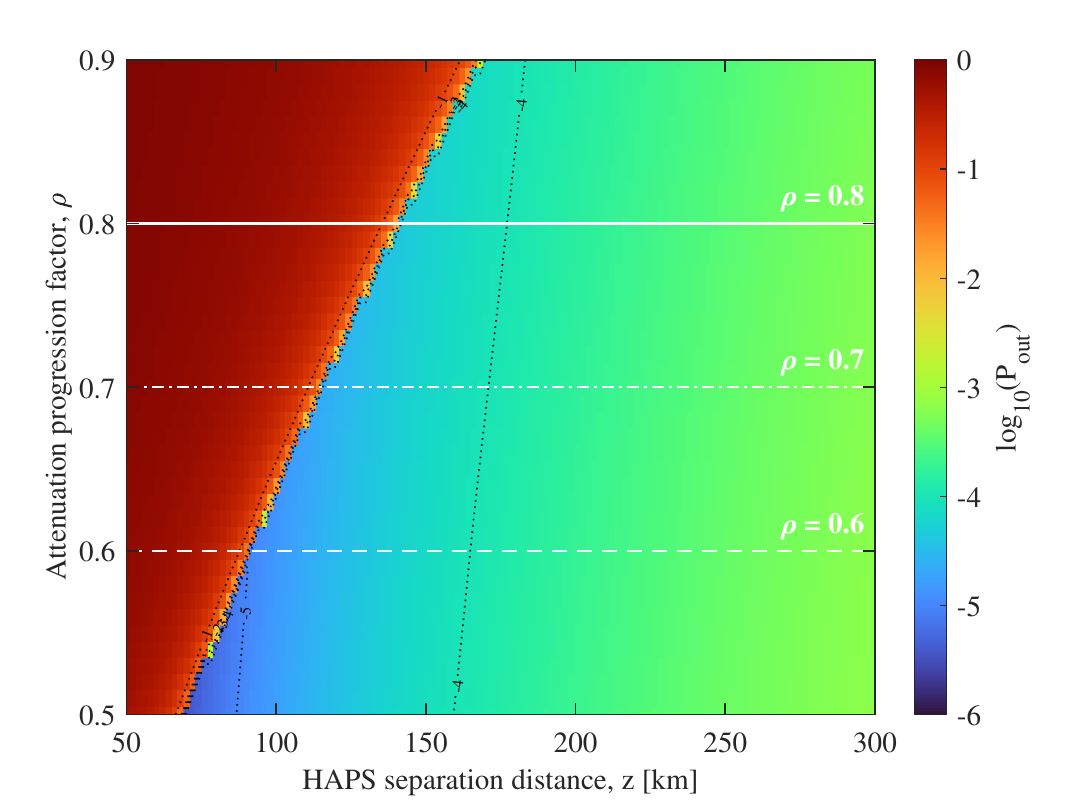}
\caption{Total system outage probability $P_{\mathrm{out}}$ for the $K=4$ square receiver as a function of HAPS separation distance $z$ and attenuation progression factor $\rho$, with $\sigma_{\theta}=10~\mu\mathrm{rad}$ and $I_{\mathrm{sat}}=5~\mathrm{mA}$.}
    \label{fig:fig3}
\end{figure}

Fig.~\ref{fig:fig3} illustrates the total system outage probability of the $K=4$ receiver
as a joint function of HAPS separation $z$ and attenuation progression factor
$\rho$. At short distances, the link is strongly saturation limited, producing
high outage, particularly for larger $\rho$, where the branches experience
weaker attenuation. Reducing $\rho$ increases the attenuation of the
higher-index branches and shifts the saturation-to-reliable-operation
transition toward shorter distances. This creates a pronounced low-outage
region, with the minimum $P_{\mathrm{out}}$ reaching approximately
$3.7\times10^{-6}$. As the separation increases beyond this region, outage
gradually rises again because beam spreading and pointing displacement reduce
the received SNR. The nearly vertical contours at larger distances further
indicate that the influence of $\rho$ becomes weaker once saturation is
relieved. Hence, $\rho$ primarily controls short-range saturation resilience,
while link distance increasingly determines performance in the longer-range
regime.

\begin{figure}[t]
    \centering
    \includegraphics[width=\columnwidth]{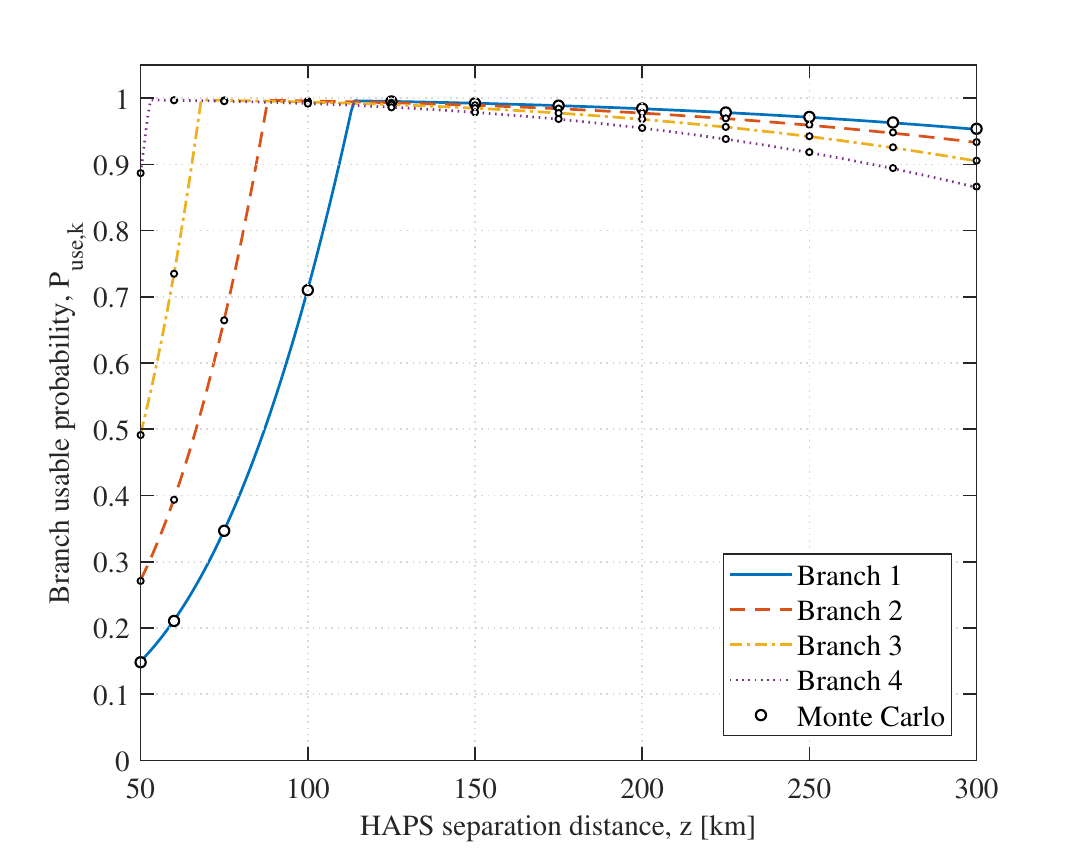}
\caption{Branch usable probability $P_{\mathrm{use},k}$ versus HAPS separation distance $z$ for the $K=4$ receiver with $\rho=0.6$, $\sigma_{\theta}=10~\mu\mathrm{rad}$, and $I_{\mathrm{sat}}=2~\mathrm{mA}$; analytical results are validated by Monte Carlo simulations.}
    \label{fig:fig4}
\end{figure}

Fig.~\ref{fig:fig4} reveals the branch-level mechanism behind the proposed dynamic-range
extension for the $K=4$ receiver. At short separations, saturation dominates,
so the more strongly attenuated branches are substantially more likely to remain
usable; at $z=50~\mathrm{km}$, $P_{\mathrm{use},k}$ increases from about
$0.15$ for Branch~1 to $0.89$ for Branch~4. As the separation increases and
saturation is relieved, all branches approach near-unity usability, with
Branch~3 reaching approximately $0.997$ around $75~\mathrm{km}$ and Branch~1
approaching $0.995$ by about $125~\mathrm{km}$. Beyond this region, the ordering
gradually reverses because stronger attenuation reduces the available SNR; at
$300~\mathrm{km}$, the usable probabilities are approximately $0.953$, $0.933$,
$0.905$, and $0.865$ for Branches~1--4, respectively. Hence, the branches
provide complementary operating regions---strongly attenuated branches protect
against short-range saturation, whereas lightly attenuated branches become
preferable as the link becomes increasingly SNR limited. The close agreement
between the analytical curves and Monte Carlo markers, with a maximum absolute
error below $1.9\times10^{-3}$, further validates the derived branch-level
usability expressions.

\section{Conclusion}
\label{sec:conclusion}

This work introduced a saturation-aware spatial multi-branch IM/DD receiver combining separated apertures, progressive optical attenuation, and maximum-SNR feasible-branch selection. Results show that the architecture substantially suppresses saturation-induced outage and extends reliable inter-HAPS operation to shorter separations. Increasing the number of branches shifts the minimum-outage point toward shorter distances and lowers outage by several orders of magnitude, while progressive attenuation controls the saturation–SNR tradeoff. Strongly attenuated branches are most effective in the saturation-limited regime, whereas lightly attenuated branches become preferable as the link becomes SNR limited. The close analytical–Monte Carlo agreement validates the derived expressions. Overall, spatially distributed attenuation provides an effective, low-complexity means of widening inter-HAPS optical-receiver dynamic range.

\bibliographystyle{IEEEtran}
\bibliography{myref}

\end{document}